# Imaging Local Effects of Voltage and Boron Doping on Spin Reversal in Antiferromagnetic Magnetoelectric Cr$_2$O$_3$ Thin Films and Devices

*Adam Erickson,*[1,*] *Syed Qamar Abbas Shah,*[2,*] *Ather Mahmood,*[2,*] *Pratyush Buragohain,*[2,$] *Ilja Fescenko,*[3] *Alexei Gruverman,*[2] *Christian Binek,*[2,&] *and Abdelghani Laraoui*[1,2,&]

[1]Department of Mechanical & Materials Engineering, University of Nebraska-Lincoln, 900 N 16th Street, W342 NH, Lincoln, Nebraska 68588, United States
[2]Department of Physics and Astronomy and the Nebraska Center for Materials and Nanoscience, University of Nebraska-Lincoln, 855 N 16th Street, Lincoln, Nebraska 68588, United States
[3]Laser Center, University of Latvia, Jelgavas St 3, Riga, LV-1004, Latvia
[*]Equal contributions
[$]Currently at Technology Research, Intel Corporation, Hillsboro, OR, USA
[&]Corresponding authors: cbinek@unl.edu , alraoui2@unl.edu

**Abstract**

Chromia (Cr$_2$O$_3$) is a magnetoelectric oxide which permits voltage-control of the antiferromagnetic (AFM) order, but it suffers technological constraints due to its low Néel Temperature ($T_N$ ~307 K) and the need of a symmetry breaking applied magnetic field to achieve reversal of the Néel vector. Recently, boron (B) doping of Cr$_2$O$_3$ films led to an increase $T_N$ > 400 K and allowed the realization of voltage magnetic-field free controlled Néel vector rotation. Here, we directly image the impact of B doping on the formation of AFM domains in Cr$_2$O$_3$ thin films and elucidate the mechanism of voltage-controlled manipulation of the spin structure using nitrogen vacancy (NV) scanning probe magnetometry. We find a stark reduction and thickness dependence of domain size in B-doped Cr$_2$O$_3$ (B:Cr$_2$O$_3$) films, explained by the increased germ density, likely associated with the B doping. By reconstructing the surface magnetization from the NV stray-field maps, we find a qualitative distinction between the undoped and B-doped Cr$_2$O$_3$ films, manifested by the histogram distribution of the AFM ordering, *i.e.*, 180° domains for pure films, and 90° domains for B:Cr$_2$O$_3$ films. Additionally, NV imaging of voltage-controlled B-doped Cr$_2$O$_3$ devices corroborate the 90° rotation of the AFM domains observed in magnetotransport measurement.

## 1. Introduction

Many remarkable discoveries in spintronics[1,2] have been achieved in the last three decades including the giant magnetoresistance effect,[4,5] current-induced magnetization switching [6,7] and ordinary and quantum spin Hall effects.[8,9] This research has accelerated with the development of magnetic tunnel junctions (MTJ)[10–12] – key devices for modern spintronic technologies, such as magnetic random-access memory (MRAM), which offer non-volatile storage but have a drawback of high writing energy. This is due to the large current densities (>10$^6$ A/cm$^2$) needed to reverse the magnetization of the ferromagnetic layer.[6,7] It would be desirable to provide non-volatility of the device state variable while being able to switch it with low power and high speed. The latter requirements can be achieved by employing antiferromagnetic (AFM) spintronics,[13] where the state variable is represented by the Néel vector. Electrical control and detection of the Néel vector in collinear AFM materials, such as Mn$_2$Au and CuMnAs,[14,15] has changed the perspective of using these materials in spintronics applications.[13] Switching the Néel vector can be performed at ultra-fast speeds, potentially reaching the THz regime (*i.e.*, nearly three orders of magnitude faster than in ferromagnets).[16] In addition, AFM materials are protected from external magnetic fields due to their zero net magnetization.[17]

Chromia ($Cr_2O_3$) is an AFM oxide, which exhibits a magnetoelectric (ME) response and hosts a surface magnetization of uncompensated spins rigidly coupled to the underlying Néel vector, Figure 1a.[17–20] When this surface magnetization is coupled to an adjacent ferromagnetic layer, voltage-controlled exchange bias (EB) emerges due to the electric 180° switching of the Néel vector and its accompanying boundary magnetization.[21] Technologically, the utility of these EB based systems is still diminished by the necessity of a magnetic field to utilize the linear magnetoelectric effect for electrical switching,[21] as well as the low Néel temperature ($T_N$) of $Cr_2O_3$ (~307 K).[22] Consequently, it is of interest to rectify these limitations in order to reap the full benefits of this materials' functionality.

First principles simulations have shown that substitutional boron doping (~3%) can increase the exchange energies between Cr spins (Figure 1b), increasing the $T_N$ and affecting magnetic anisotropy.[23] Very recently, AFM order up to 400 K and above was demonstrated experimentally in B:$Cr_2O_3$ films.[16,24] Furthermore, from magnetotransport measurements, B:$Cr_2O_3$ was predicted to exhibit a modified bistability of the entrained Néel vector, which supports both in-plane and out-of-plane AFM ordering, and leads to a 90° rotation of the Néel vector when switched by an applied electric field.[16] The underlying mechanism was attributed to polar nanoregions (PNRs), which possess an electrical dipole moment and can be aligned by an electric field inducing macroscopic polarization. The polarization is accompanied by piezoelectric straining which, *via* magnetoelastic coupling, changes the magnetic anisotropy and thus leads to a realignment of the magnetic easy axis.[16] However, there has been no direct evidence of this phenomenon, in which the magnetoresistance can be correlated with the 90° rotation of the surface magnetization.

Magnetic microscopy based on nitrogen-vacancy (NV) centers in diamond has become a prevalent tool to study antiferromagnets due to the high spatial resolution, high sensitivity to magnetic field, and lack of magnetic back-action.[25–31] In this work, we use NV scanning probe magnetometry (NV-SPM)[32,33] to study the effect of boron doping and thickness on the emergent AFM domain structure (see Figures 1a and 1b) in $Cr_2O_3$ films (50 – 300 nm thick) grown on $Al_2O_3$ substrates with $V_2O_3$ bottom electrode. By analyzing the reconstructed surface magnetic moment density maps, we find a trend for AFM domain size with respect to the film thickness, which is interpreted phenomenologically through a simulated magnetic phase transition following the tenets of Avrami kinetics of crystalline solids.[34] Furthermore, we demonstrate 90° rotation of the Néel vector in B:$Cr_2O_3$, contrary to pure films with 180° rotation. This is the first direct imaging of such behavior and supports the proposed mechanism of earlier magnetotransport measurements.[16,21] Finally, a spatially resolved voltage-induced switching effect of the Néel vector in B:$Cr_2O_3$ is demonstrated. Our study provides useful details for the implementation of B:$Cr_2O_3$ in functional AFM spintronics devices.

## 2. Results and discussion:

### 2.1. Characterization of pure and B doped $Cr_2O_3$ films

The structural properties of $Cr_2O_3$ and B:$Cr_2O_3$ of various thicknesses (50 – 300 nm) grown on $V_2O_3$ (20 nm)/sapphire ($Al_2O_3$) substrates were characterized by using X-ray diffraction (XRD). Representative XRD spectra for 150 nm $Cr_2O_3$ film and B:$Cr_2O_3$ are shown in Figures 1c and 1d, respectively, with clearly distinguished peaks corresponding to the (006) and (0012) planes of the V$_2$O$_3$, Cr$_2$O$_3$, and Al$_2$O$_3$ layers. XRD measurements did not reveal significant structural changes between the pure and B-doped Cr$_2$O$_3$ films, which is likely due to the small B doping percentage (~3%).[16,35,36] Atomic force microscopy imaging on the same Cr$_2$O$_3$ (Figure 1e) and B:Cr$_2$O$_3$ (Figure 1f) films showed low surface roughness with a root mean square (rms) of ~0.292 nm. In previous theoretical[37,38] and experimental studies,[20,28] the surface magnetization was found to be largely roughness insensitive, *i.e.*, the surface magnetization represents the underlying order parameter despite small changes in the topography. To investigate the effect of thickness and B doping on the resulting AFM domains in Cr$_2$O$_3$, a set of pure and boron doped films with thicknesses of 50 nm, 100 nm, 150 nm, 200 nm, and 300 nm were grown on V$_2$O$_3$/Al$_2$O$_3$ substrates. The films were characterized by XRD (Figure S1.1, in Supporting Information (SI)) to verify crystallinity, and atomic force microscopy to ensure low surface roughness (Figure S1.2, SI), see the Supporting Information (SI) note S1 for further details.

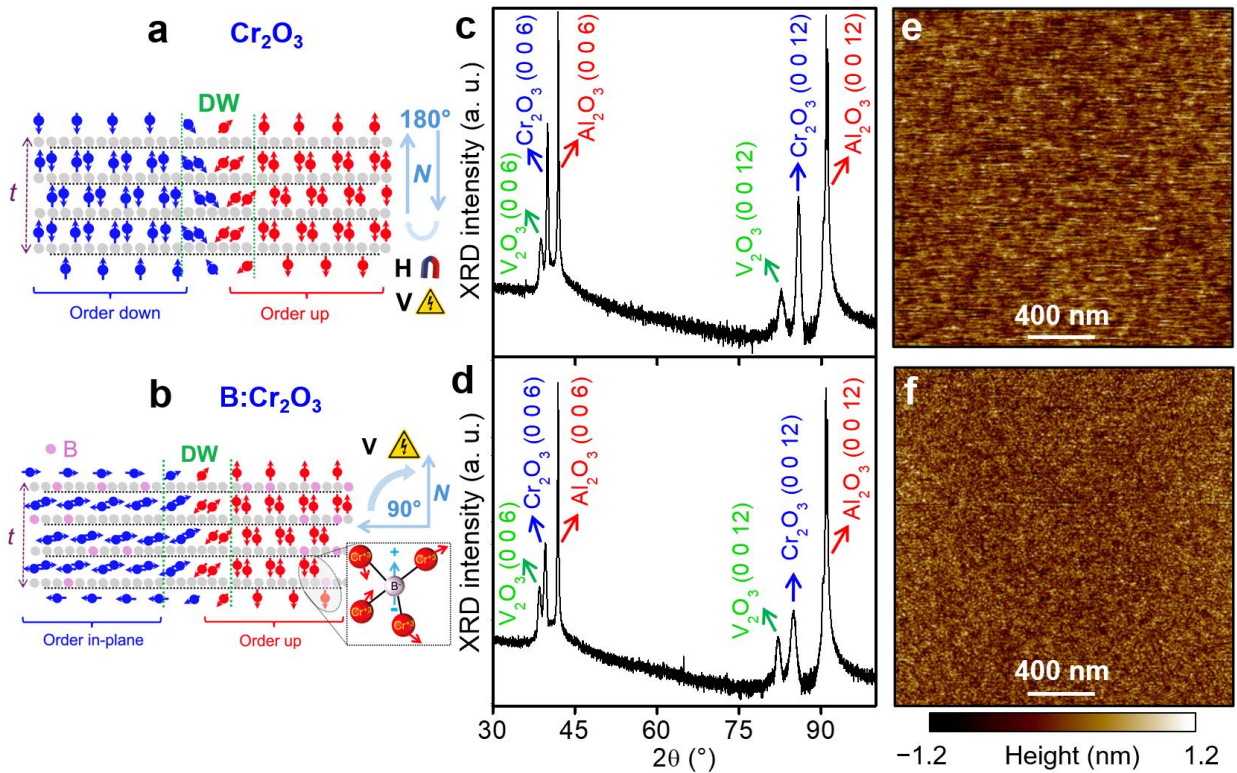

**Figure 1. Characterization of pure and boron doped Cr$_2$O$_3$ films.** (a) Expected collinear antiferromagnetic spin ordering for undoped Cr$_2$O$_3$, with uncompensated surface termination giving rise to a parasitic magnetic moment. The isothermal 180º rotation of the Néel vector requires the application of a supercritical product of electric and magnetic field. (b) Predicted spin configuration of B:Cr$_2$O$_3$ with 90º between Néel vector stability. (Inset) The presence of polar nano-regions allows for voltage-only switching between antiferromagnetic states via strain coupling to the magnetic order. (c,d) XRD spectrum of pure (c) and B doped (d) Cr$_2$O$_3$ films. (e,f) Topographic images acquired on 150 nm pure (e) and B doped (f) Cr$_2$O$_3$ films.

## 2.2. NV magnetometry of AFM domains in pure and B doped Cr$_2$O$_3$ films

Imaging the magnetic states of AFM materials via traditional techniques, such as magnetic force microscopy (MFM), is highly challenging due to their zero net magnetization. In MFM, apparent contrast due to magnetic forces is often affected by other long-range forces associated, for instance, with surface charges that make it hard to quantitatively interpret the measured magnetic signals.[39] Only a handful of techniques are used to study AFM materials, such as X-Ray Magnetic Circular Dichroism (XMCD) and X-Ray Photo Emission Electron Microscopy (PEEM).[39] However, these experimental tools tend to suffer from low spatial resolution (> 50 nm in PEEM[40]) or are expensive and require complex equipment. Recently, an alternative technique for measuring magnetic fields at the nanometer scale based on optical detection of the electron spin resonances of NV centers in diamond has emerged.[41–45] Negatively charged NV centers, comprised of a substitutional nitrogen atom adjacent to a vacancy site, are bright, photostable emitters that exhibit optically detected magnetic resonance (ODMR).

Magnetic images of the magnetic stray field produced at the AFM domain boundaries were obtained by NV-SPM (Figure 2a). Here, the scanning diamond probe is optically addressed with a 532 nm laser, and the spin dependent fluorescence signal (650 – 750 nm) is collected in the epifluorescence configuration.[28,46,47] The application of a bias magnetic field (40 Oe) does not influence the AFM domains of the film, and allows for sign sensitive measurement of the stray field.[28] The NV measured value of the stray field at a distance $z$ can be described by $B_{\text{str}} = \boldsymbol{B} \cdot \hat{u}_{\text{NV}}$, where $B_{\text{str}}$ is the projection of the vector field, $\boldsymbol{B}$, on to the sensing axis, defined by the NV symmetry axis, and represented by the unit vector $\hat{u}_{\text{NV}}$. For this study, a (100) grown diamond probe was used with the NV axis along [111] orientation. Figure 2b displays $B_{\text{str}}$ map of 150 nm $Cr_2O_3$ in which the regions of the AFM domains are defined by sizeable demagnetization fields present at the edge of two surface magnetization orientations. By using the values of the calibrated NV sensor geometry ($z$ = 50 nm, $\theta$ = 48°, and $\phi$ = 92°), the surface magnetic density $\sigma_z$ can be reverse propagated [26,43] from $B_{\text{str}}$ maps. The acquired $\sigma_z$ image on 150-nm-thick undoped $Cr_2O_3$ film confirms the presence of homogeneously magnetized domains, as signaled by areas of positive (red) or negative (blue) $\sigma_z$ with sharp domain walls, Figure 2c.

A similar approach was used to extract $\sigma_z$ map (Figure 2e) from $B_{\text{str}}$ map (Figure 2d) in 150 nm thick B:$Cr_2O_3$. There is a visible difference in the size of the AFM domains present in B:$Cr_2O_3$ films in comparison to the pure films (note that the scale bars in Figures 2d and 2e are 0.5 µm *vs* 1 um in Figures 2b and 2c). Compared to the pure films, the AFM domains appear blurrier due to the higher frequency components of the image being filtered by the Hanning window (see SI Note 2) which is implemented to enforce the resolution defined by the distance $z$ between the NV sensor and the chromia surface.[26,28] The sizes of individual AFM domains are characterized by a mean radius of the amorphous boundary magnetization (Figure S2.2, in SI) and plotted as a histogram for both pure and doped films (Figure 2f). The distributions of domain sizes are fitted with a lognormal function. The pure $Cr_2O_3$ films are characterized by a much broader distribution, and host larger domains in the hundreds of nanometers. By comparison, the distribution of domain sizes for the B:$Cr_2O_3$ film shows a tighter grouping with a significantly smaller average radius (<100 nm).

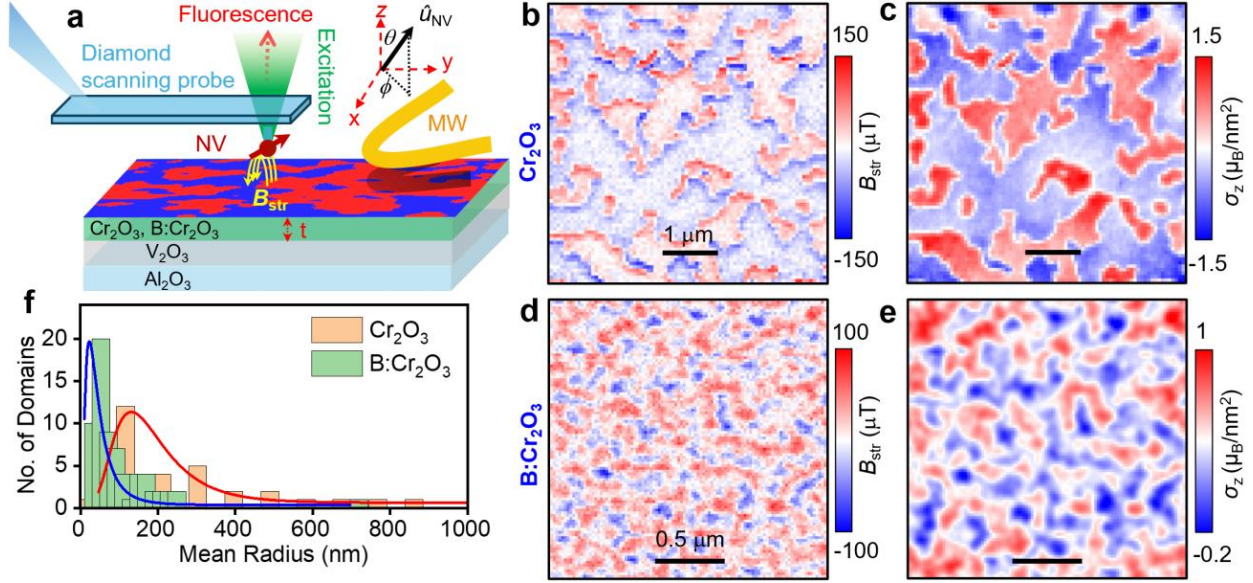

**Figure 2. Effect of boron doping on AFM domain size.** (a) Schematic of films heterostructure and sketch of NV-SPM measurement configuration. Stray magnetic fields (yellow lines) are measured via the ODMR of the NV center implanted at the apex of the diamond pillar. The coordinate system defines the sensing axis, $\hat{u}_{NV}$, onto which the measured vector field is projected. (b) $B_{str}$ map of 150 nm pure $Cr_2O_3$. (c) Reconstructed $\sigma_z$ map calculated from $B_{str}$ map in (b). (d) $B_{str}$ map of 150 nm B-$Cr_2O_3$. (e) Reconstructed $\sigma_z$ map calculated from $B_{str}$ map in (d). (f) Histogram of the mean radius of individual domain boundaries as estimated through binarization and fitted by the lognormal distribution.

## 2.3. Effect of thickness and B doping on AFM domains in $Cr_2O_3$ films

To investigate further the effect of B doping on the size of the AFM domains, NV-SPM measurements were performed on a series of pure and boron doped $Cr_2O_3$ films, which were grown at different thicknesses. Figure 3a shows the magnetic moment density, $\sigma_z$, maps of 50 nm, 150 nm, 200 nm, and 300 nm pure $Cr_2O_3$ films, characterized by large AFM domains of well-defined magnetization regardless of the thickness of the film. Some of the $Cr_2O_3$ films (*e.g.*, 200 nm) exhibit a larger number of small domains, which may be related to the subtle fluctuations in the quality of structure from film to film. Figure 3b shows the $\sigma_z$ maps of 50 nm, 100 nm, 150 nm, and 200 nm of B:$Cr_2O_3$ films. As with the case of 150 nm thin films discussed above (Figure 2f), there is clear correlation with B doping and film thickness on the domain size of $Cr_2O_3$ films.

To visualize the empirical relation between the thickness and the AFM ordering, we performed size analysis of binarized magnetization maps and plotting the average mean radius, $\bar{R}_m$, for the population (Figure 3c). Plotted as a shaded area, the interquartile range (IQR) is the difference of the median of the upper and lower half of the logarithmic data set, which is much higher for the pure films due to the tendency to form domains that are orders of magnitude wider than the thickness of the film itself. The B:$Cr_2O_3$ films exhibit a larger correlation between the resulting AFM domain size and the thickness of the film. It has been shown via micromagnetic simulations that the exchange energies associated with the grain boundaries can serve to drive the formation of much smaller domains when dominated by ferromagnetic coupling.[48] Conversely, when the antiferromagnetic exchange is dominant, large domains of homogeneous order parameter emerge. Furthermore, the uniaxial anisotropy orientation can also be a determinant of the resulting domain structure. In our earlier work, which predicted the effect of the boron doping on the density of states of $Cr_2O_3$, the increase of $T_N$ was mostly attributed to the modification of the local ferromagnetic exchange energy due to the hybridization geometry.[23,49] While the notion that the enhancement of the exchange energy due to boron doping[35] results in more fragmented AFM domains is in accordance with micromagnetics,[48] it does not encapsulate the thickness dependence of the resulting domain structure. Furthermore, the magnetic phase transition from paramagnetic to antiferromagnetic when cooling through $T_N$ depends on the density of nucleation sites,[20,28] which are attributed to defects primarily located at the interfaces and grain boundaries.

In order to investigate the role of the Néel transition on the final domain configuration, we modeled the paramagnetic to antiferromagnetic phase transition using the principles of Avrami phase kinetics.[34] The results of Monte-Carlo Avrami simulations (see Experimental Section 4) are summarized in Figure 3d, where $\bar{R}_m$ is plotted for different thicknesses of films containing a higher areal germ density (AGD) of 1% or a lower AGD of 0.2%. Linear fits of the trends in the domain size *vs* the thickness reveal a greater slope for the films with AGD of 1%. Linear fits of the trends in the domain size *vs* the thickness reveal a greater slope for the films with AGD of 1%. The inset of Figure 3d displays the experimental data set which demonstrates similar scaling factors when comparing the pure and doped films. The shaded region again displays the IQR for the distribution associated with each thickness. Such distributional widths also capture the large variation of resulting sizes as were measured experimentally. The top layers of the simulations, akin to the surface magnetization, were plotted in Figures 3e and 3f. The spatial morphology of the resulting domain structure, while qualitatively similar, lacks the influences of magnetostatic energy minimization.

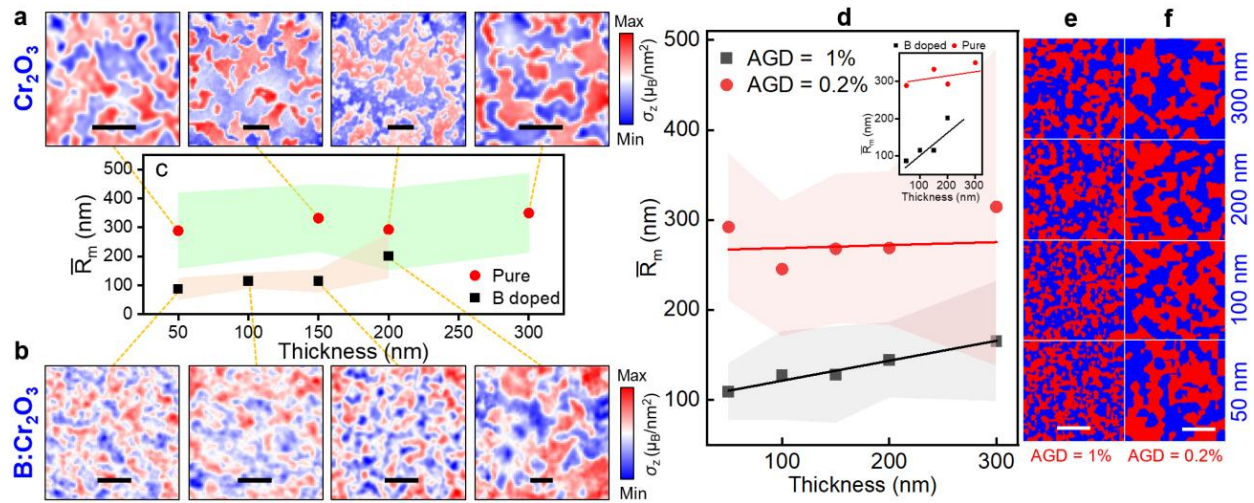

**Figure 3. Thickness dependence of AFM domains in $Cr_2O_3$ and $B:Cr_2O_3$.** Representative $\sigma_z$ reconstruction for several thicknesses of pure (a) (scale bar 1 μm) and B-doped (b) (scale bar 500 nm) chromia. (c) $\bar{R}_m$ plotted as a function of the film thickness. Shaded regions represent the IQR of the size distribution associated with each thickness. (d) Average $R_m$ as a function of the simulated thickness for different areal germ densities. The shaded area represents the IQR of the size distribution. Inset in (d) Experimentally obtained size vs thickness for reference. Resulting domain structure when simulating AGD of 1% (e) and 0.2% (f) for various thicknesses. The scale bar is 1 μm.

## 2.4. Direct visualization of the AFM domains orientation in pure and B-doped $Cr_2O_3$ films

Along with the analysis of size, directional analyses of the reconstructed surface magnetic moment density were carried out for $Cr_2O_3$ films, which hosted relatively large magnetization patterns, ensuring high fidelity. We compared the reconstructions of $\sigma_z$ for 200 nm $Cr_2O_3$ (Figure 4b) and B:$Cr_2O_3$ (Figure 4e) based on the measured stray field maps for each film shown in Figures 4a and 4d, respectively. When plotted with a symmetric color scale, the $\sigma_z$ map of the pure $Cr_2O_3$ film consists of roughly equal regions of positive and negative moment density whereas the $\sigma_z$ map of B:$Cr_2O_3$ film shows mostly red regions (up to 2.5 $\mu_B/nm^2$) with other domains of white or faint blue indicating a near zero value of $\sigma_z$.

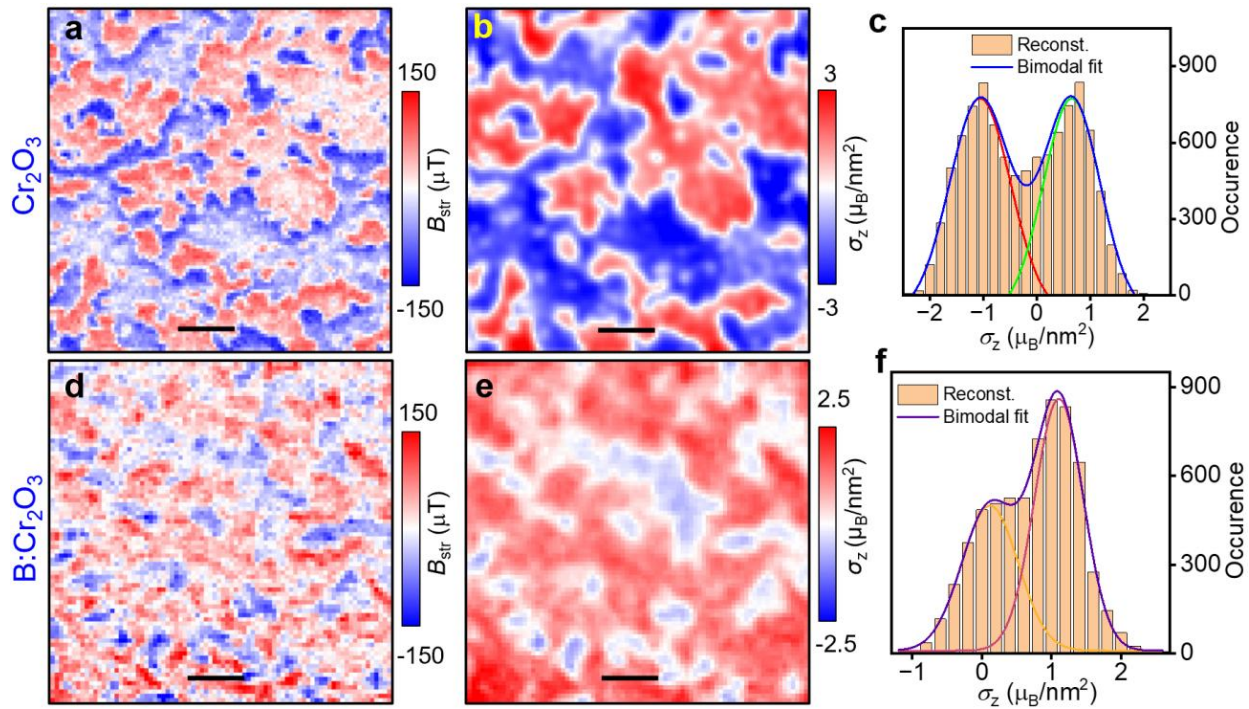

**Figure 4. Effect of B doping on the bistability of the Néel vector.** $B_{str}$ map (a) and $\sigma_z$ map (b) of 200 nm $Cr_2O_3$ film. (c) Histogram of $\sigma_z$ for image (b), displaying near symmetry of distinct moment populations about zero. $B_{str}$ map (d) and $\sigma_z$ map (e) of 200 nm B-$Cr_2O_3$. (f) Histogram of $\sigma_z$ for image (e), revealing the coexistence of an out-of-plane magnetization as well as a population with zero average, suggesting in-plane moment. The scale bar for all images is 500 nm.

The effect of doping is more easily interpreted from the histogram representation of $\sigma_z$ maps. For $Cr_2O_3$ (Figure 4c), the histogram is centered around zero, which is an established property of $Cr_2O_3$ arising from the antiparallel alignment of the stable Néel vector.[26,28] This fits well by a bimodal distribution, which describes the two stable populations of AFM order. The histogram of the reconstructed $\sigma_z$ for the B:$Cr_2O_3$ (Figure 4f) is shifted positively in the moment axis compared to the pure distribution. When fitted by the bimodal distribution, two populations are identified. There is one population present at a mean value of ~1 $\mu_B/nm^2$ like in the pure films, and one which has a mean value of zero not observed in undoped $Cr_2O_3$. This implies that the latter AFM domain population can be described either by a lack of antiferromagnetic ordering and thus the loss of the surface magnetization altogether,[27] or by the rotation of the Néel vector to an in-plane orientation (90° rotation),[16] such that the projection of the moment density onto the z-axis would yield zero on average. The coexistence of magnetically long range ordered and disordered regions in a homogeneous single-phase material would be a thermodynamic curiosity which is highly unlikely to occur. In contrast, the presence of a canted Néel vector in B:Cr$_2$O$_3$ with 90° rotation was earlier proposed as a model to explain the asymmetric magnetic field free voltage-induced switching results from the magnetotransport measurements.[16]

### 2.5. Magnetic field free switching of the Néel vector

Voltage-controlled switching of the Néel vector enables dissipationless control of remnant magnetic states for nonvolatile data storage.[21] Due to the symmetry-protected boundary magnetization at the interface of ME AFMs,[20,50] single phase ME are very promising for voltage-controlled spintronics.[13] In sharp contrast to spin-torque transfer MRAM, which is energy expensive and limited to a nanosecond switching speed, the ME interface can lead to energy-efficient (attojoule) and ultrafast (THz) device operation.[51] In previous studies, a robust and deterministic switching of boundary magnetization in Cr$_2$O$_3$ was demonstrated well above RT using voltage while a symmetry breaking magnetic field was applied to activate the magnetoelectric energy contribution to the free energy which is proportional to the product of electric and magnetic field and thus zero in zero magnetic field.[16] It was also found that B:Cr$_2$O$_3$ film adjacent to a non-magnetic patterned Pt Hall bar exhibits a hitherto little explored switching mechanism that eliminates the need for an applied magnetic field. The Hall bar enables readout of the AFM interface (boundary) magnetization through detection of a transverse voltage signal, $V_{xy}$, in response to in-plane current density $J_z$ (see SI Note 3). The Hall-like signal $V_{xy}$ is widely believed to originate from spin Hall magnetoresistance and from the anomalous Hall effect (AHE) caused by the proximity induced in the Pt Hall bar by the exchange field of the boundary magnetization. The application of a voltage, $V_G$, across the film switches the film between distinct non-volatile AFM states associated with $V_{xy} \approx 0$ and $V_{xy} > -10$ mV and suggests 90° rotation of the Néel vector in sharp contrast to Hall signals observed for 180° switching in undoped Cr$_2$O$_3$ films. Rotation of the Néel vector by 90° is consistent with the non-broken time reversal symmetry by an electric field.[16]

To further understand the switching mechanisms in B:Cr$_2$O$_3$, we fabricated few-micrometer square capacitor structures by milling top electrodes from the deposited Pt layer using focused ion beam (FIB). Figure 5b shows a scanning electron microscope (SEM) image of the Pt/B:Cr$_2$O$_3$/V$_2$O$_3$ devices. The electric field was applied through the capacitor structure by a biased conductive atomic force microscope probe in direct contact with the top electrode (Figure 5a). By fabricating top electrodes of different sizes, we were able to iteratively determine the voltage application limits corresponding to the leakage breakdown of the device. Details regarding the device fabrication and characterization are found in SI Note 3. Prior to the application of a voltage pulse, capacitor structures were imaged by NV-SPM, during which the NV photoluminescence (PL, Figure 5c) and ODMR were captured simultaneously. Due to the milling of the thin Pt layer by FIB, the B:Cr$_2$O$_3$ layer is exposed, which generates a clear quenching contrast in the PL channel and serves to identify the boundary of the top electrode. The reconstructed $\sigma_z$ map for the pristine condition is shown in Figure 5d. A histogram of $\sigma_z$ values for the dashed area is presented in Figure 5e, which corresponds to the boundaries of the FIB milled Pt. This distribution also provides independent confirmation of the asymmetrically distributed moment as in Figure 4f. Only the area directly under the top electrode in the capacitor structure is considered for a comparison of the switched volume, as surrounding locations will have experienced an *E*-field which deviates from out of plane. Following the application of a -11 V, 100 µs pulse, the same region was imaged again (Figure 5f, 5g). The $\sigma_z$ histogram after voltage application (Figure 5h) reveals a shift in the relative populations, favoring the out of plane component, suggesting 90° rotation of the Néel vector. After the application of the voltage pulse, it can also be seen that regions outside of the capacitor structure were also switched, due to the surrounding inhomogeneous *E*-field.[16] There are also regions within the area of the top electrode which did not switch, suggesting that there may be pinned domain structures, induced from FIB residue, which do not participate in switching.[28]

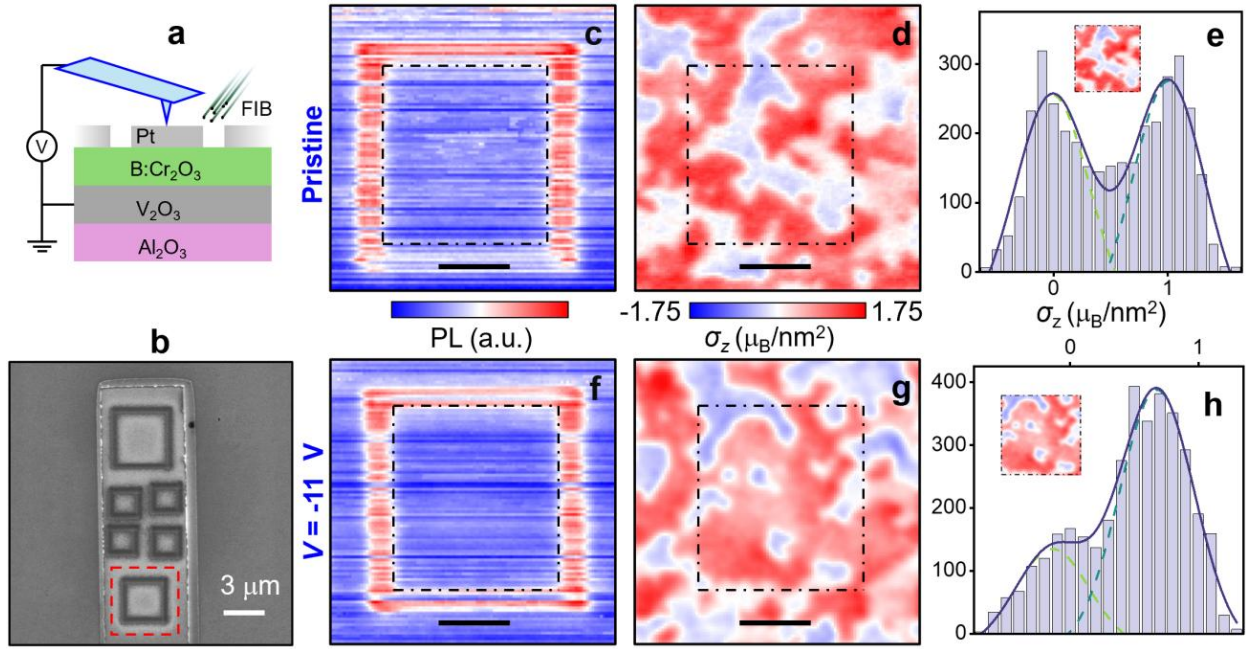

**Figure 5. Reorientation of the AFM ordering upon application of electric field.** (a) Schematic for the application of voltage pulses through FIB fabricated capacitor structure. The voltage is applied via a conductive atomic force microscope probe. (b) SEM image of the FIB milled capacitor structures. The imaged capacitor is denoted by the red dashed square. (c) PL image obtained simultaneously during ODMR imaging, revealing the outline of the capacitor due to the lack of quenching associated with the etched platinum region. Reconstructed $\sigma_z$ map (d) and histogram (e) prior to the application of electric field. PL (f), reconstructed $\sigma_z$ map (g), and histogram (h) of $\sigma_z$ after the application of -11 V, respectively. The scale bar in (c,d,f,g) is 1 µm.

## 3. Conclusion

To conclude, we have studied the effects of boron doping on the formation of antiferromagnetic domains in $Cr_2O_3$ films using NV magnetic microscopy. A significant reduction and thickness dependence of the AFM domain size upon doping with boron was found and simulated by the paramagnetic to AFM phase transition based on Avrami theory. From the reconstructed surface magnetic moment density maps from NV stray-field measurements, we observe a 90º rotation between the stable Néel vector states in $B:Cr_2O_3$. The result gives a strong corroboration to the model for the asymmetric spin Hall magnetoresistance measurements, explained by voltage induced 90º rotation of the Néel vector.[16] Finally, the demonstration of the reorientation of the surface magnetization upon application of an electrical field/voltage provides even more salient evidence of this behavior, although this is just a preliminary result toward the development of reliable AFM spintronics devices. [13]

Indeed, further studies, which directly correlate modulation of $V_{xy}$ to the component resolved magnetization in the same device, will help reveal the exact mechanisms of functionality. Furthermore, initial piezoresponse force microscopy (PFM) measurements showed that application of voltage can induce a polarization manifested via observation of the PFM butterfly hysteresis loops in B:Cr$_2$O$_3$/V$_2$O$_3$ devices,[16] not observed in the undoped Cr$_2$O$_3$ films. A critical hypothesis to be tested is that the induced polarization may couple to the AFM order. The induced polarization is explained by electric field-induced alignment of dipole moments in PNR regions, which are not yet detected experimentally. Future systematic PFM, transmission electron microscopy, magnetotransport, and NV magnetometry measurements on B:Cr$_2$O$_3$/V$_2$O$_3$ devices may reveal the presence of PNRs, understand the dynamics of induced polarization, and provide a description of the polarization coupling to the Néel vector and surface magnetization.

## 4. Experimental Section

*Film growth*: Both Cr$_2$O$_3$ and B:Cr$_2$O$_3$ of various thicknesses (50 – 300 nm) were grown on sapphire single crystalline substrates (*c*-plane) by using the pulsed laser deposition (PLD) in an ultrahigh vacuum with a base pressure of $\approx 5.0 \times 10^{-9}$ Torr ($\approx 7.0 \times 10^{-7}$ Pa). The sapphire substrates were cleaned using a modified Radio Corporation of America protocol.[52] A (0001) oriented V$_2$O$_3$ film directly grown on sapphire substrate serves as a bottom electrode, followed by the growth of Cr$_2$O$_3$ (and B:Cr$_2$O$_3$) films. The substrates were heated to 830 ºC during V$_2$O$_3$ deposition. A KrF excimer laser with pulse energies of 200 mJ, a spot size of $\approx$ 6 mm$^2$, and a pulse width of 20 ns (at a repetition rate of 10 Hz) was used to abate V$_2$O$_3$ target. The target to substrate distance was kept at $\approx$ 9 cm and substrate rotation rate was 4 rpm. The growth rate of V$_2$O$_3$ was $\approx$ 0.017 nm/s, and the thickness was maintained at 20 nm. The Cr$_2$O$_3$ (and B:Cr$_2$O$_3$) films were then grown at 800 ºC on the resulting substrates (described above) using PLD. The pulse energy of 190 mJ and a frequency of 10 Hz is used to ablate the Cr$_2$O$_3$ target. The B:Cr$_2$O$_3$ films deposition takes place in the presence of a decaborane $(B_{10}H_{14})$ background gas with a vapor pressure of $1.0 \times 10^{-4}$ Pa. Various thicknesses of Cr$_2$O$_3$ and B:Cr$_2$O$_3$ films were grown at a growth rate of $\approx$ 0.025 nm/s.

*NV Magnetometry Measurement:* We used a home-built NV-SPM hosting tuning-fork-based atomic force microscope and optical confocal microscope.[28] The unique capabilities of the system allow imaging of various magnetic materials at variable temperature (290 – 400 K) and magnetic fields up to 30 mT using permanent magnets along [111] NV orientation in (100) diamond. The challenge of integrating the tip in the tight space between the objective and magnetic sample was solved by using a long working distance (6 mm) Mitutoyo objective (0.7 NA). The self-sensing SPM modality of the tuning fork-based feedback negates the need for laser excitation, which would conflict with NV fluorescence measurements. The tuning fork is attached to Attocube stack with 3-axis positioning for coarse approach and fine positioning. An additional piezo-based *xyz* scanner (Npoint) provides a scan range of 100 μm × 100 μm × 20 μm. The NV-SPM is integrated with a freestanding microwave antenna (MW) to measure NV spin transitions. We perform NV ODMR measurements by sweeping the MW frequency along the NV transitions $|m_S = 0\rangle$ to $|m_S = \pm 1\rangle$ at a given applied magnetic field $B_{app}$. The Hamiltonian of the system is: [42]

$$H = DS_z^2 - \gamma_{NV}\big(S_x(B_{app,x} + B_{str,x}) + S_y(B_{app,y} + B_{str,y}) + S_z(B_{app,z} + B_{str,z})\big),$$

where $\gamma_{NV} = 28$ GHz/T is the gyromagnetic ratio of the electron spin. The second term of the Hamiltonian is the Zeeman splitting term. By increasing/decreasing the amplitude of $B_{app}$, the resonance of the $|m_S = 0\rangle$ to $|m_S = -1\rangle$ and $|m_S = 0\rangle$ to $|m_S = +1\rangle$ transition peaks shift to higher/lower frequencies. By monitoring the NV fluorescence increase/decrease, we can measure the magnetic field strength generated by the magnet and the additional local stray field $B_{str}$ generated from scanning across the AFM spin textures in pure and B doped $Cr_2O_3$ films. This is the basis of DC magnetic sensing scheme of NV magnetometry.[41]

*Avrami Modeling of Magnetic Domains:* We define a three-dimensional matrix to serve as the simulation meshwork of our thin film system undergoing the Néel transition. The simulation size for this work is $300 \times 300 \times (t/10)$, where $t$ is the thickness of the film, such that the voxel size is 1000 $nm^3$. Germs representing the nucleation centers of two stable antiferromagnetic orders are randomly distributed through the paramagnetic medium at a prescribed areal germ density. For each time step of an iteration sequence, the germs grow with spherical wavefronts, and growth stops at points where domains of opposite ordering contact one another. The resulting surface magnetization states are then subjected to the same size analysis performed on the experimentally obtained moment density reconstructions.

**Acknowledgments**

This work is supported by the National Science Foundation/EPSCoR RII Track-1: Emergent Quantum Materials and Technologies (EQUATE), Award OIA-2044049. I. F. acknowledges support from the Latvian Quantum Initiative under European Union Recovery and Resilience Facility project no. 2.3.1.1.i.0/1/22/I/CFLA/001. The research was performed in part in the Nebraska Nanoscale Facility: National Nanotechnology Coordinated Infrastructure and the Nebraska Center for Materials and Nanoscience (and/or NERCF), supported by the National Science Foundation under Award ECCS: 2025298, and the Nebraska Research Initiative.

**Competing financial interests**

The authors declare that they have no competing financial interests.

**Author Contributions**

A.E. performed NV-SPM, analyzed the data, and created the code for the simulation; A.M and S.Q.A.S grew the pure and B doped $Cr_2O_3$ films and performed topography and XRD measurements; A.M. fabricated the hall bar devices and performed electrical switching experiments; P.B. under the supervision of A.G. performed voltage control *via* conducting SPM probe. I.F. created the Mathematica code for recovering vector stray-field components. C.B. and A.L. designed the experiments and supervised the project. A.E. wrote the paper with support from A.L. and contributions and feedback from all authors.

**Data Availability**

The data that support the findings of this study are available from the corresponding author upon reasonable request.

**Keywords**

nitrogen-vacancy, chromia, boron, magnetoelectricity, antiferromagnet, Néel vector orientation, voltage induced switching

# Supplementary Information

## Supplementary Note S1. Characterization of pure and boron doped $Cr_2O_3$ films

Additional X ray diffraction (XRD) analysis was performed on pure $Cr_2O_3$, and boron doped $Cr_2O_3$ (B: $Cr_2O_3$) films with thicknesses of 50 nm, 100 nm, 200 nm, and 300 nm grown on $V_2O_3$/sapphire ($Al_2O_3$) substrates. In each case, the films exhibited the expected XRD peak positions for the $Cr_2O_3$ film, as well as the $Al_2O_3$ and $V_2O_3$ (see Figure S1.1a-c). The XRD spectra of B:$Cr_2O_3$ films (Fig S1.1d-f) did not show any noticeable effects as a result of the low (~ 3 %) boron doping, compared to the undoped film spectra.

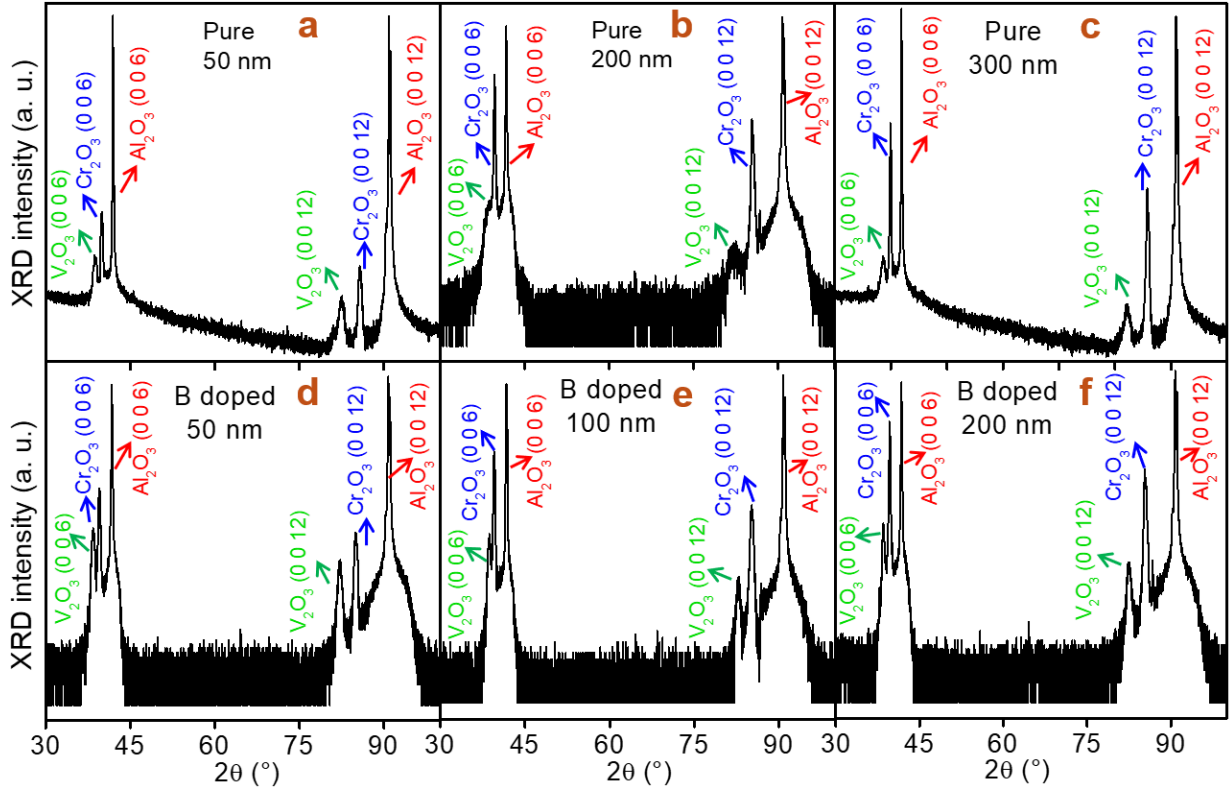

**Figure S1.1.** XRD spectra of 50 nm (a) 100nm (b) and 300 nm (c) $Cr_2O_3$. (d-f) XRD spectra of the doped films of thicknesses 50 nm (d) 100 nm (e) and 200 nm (f).

In addition to the verification of structure with XRD, atomic force microscopy measurements were performed to guarantee low film roughness prior to characterization by nitrogen-vacancy scanning probe microscopy (NV-SPM). The topography scans for pure of 50 nm, 200 nm, and 300 nm $Cr_2O_3$ films are shown in Figures S1 2a-c. The topography images of 50 nm, 100 nm, and 200 nm B:$Cr_2O_3$ films are depicted in Figures S1.2d-f. In each film, the RMS roughness, $R_q$, was found to be less than 300 pm.

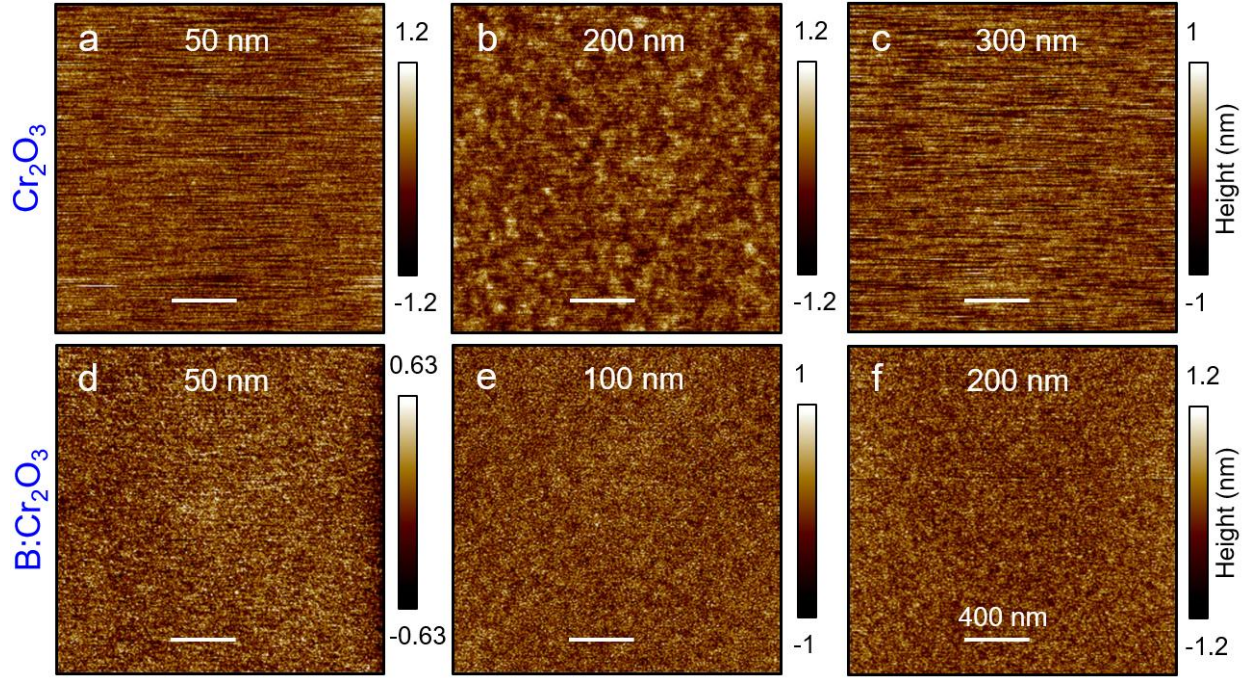

**Figure S1.2.** Surface roughness characterization of the measured films. Topography images by atomic force microscopy for 50 nm (a), 200 nm (b) and 300 nm (c) $Cr_2O_3$ films. Topography images of 50 nm (d) 100 nm (e) and 200 nm (f) $B:Cr_2O_3$ films.

## Supplementary Note 2: NV measurements and analysis protocols

Quantitative NV magnetic stray field $B_{str}$ images were obtained using a home-built NV-SPM, as detailed in reference [28]. Figure S2.1a displays a photoluminescence (PL) image of single NV center. Figure S2.1b shows a representative optical detected magnetic resonance (ODMR) peak corresponding to the $|m_s = 0\rangle$ to $|m_s = +1\rangle$ NV spin transition. In some cases, particularly magnetic images which were subjected to detailed magnetization reconstruction analysis, the fully quantitative stray field maps were obtained by tracking the ODMR center frequency at each pixel of the scan area. In other cases, to speed up the acquisition time, a dual iso-B approach was taken, which only measures the PL at two applied MW frequencies (Figure S2.1b).[53] The construction of the sign sensitive image is shown in Figures S2.1c-d, in which the normalized PL captured at a microwave (MW) frequency $f_1$ is subtracted from that of MW $f_2$. When comparing images taken by either full ODMR tracking or dual iso-B, the typical magnetic stray field $B_{str}$ induced Zeeman shift (< 5 MHz) are such that estimations are enabled by a linear approximation.[26,43] The improved minimum detectable static magnetic field in the ideal photon-shot-noise limit is given by:[44,54] $B_{min} \cong 4\, \Gamma\, (3\sqrt{3}\, \gamma_{NV}\, C)^{-1}\, (I_0\, t)^{-1}$, where $\Gamma$ is the full-width-at-half-maximum linewidth of the ODMR peak, $C$ is the ODMR peak contrast, $I_0$ is the NV PL rate, and $t$ is the measurements time.[55] By using the parameters of the NV measurements in Fig. S2.1, ($I_0$ = 500k counts/s, $\Gamma$ = 7.78 MHz, C = 0.15) we found $B_{min} = 5\ \mu T$ for $t$ = 1s.

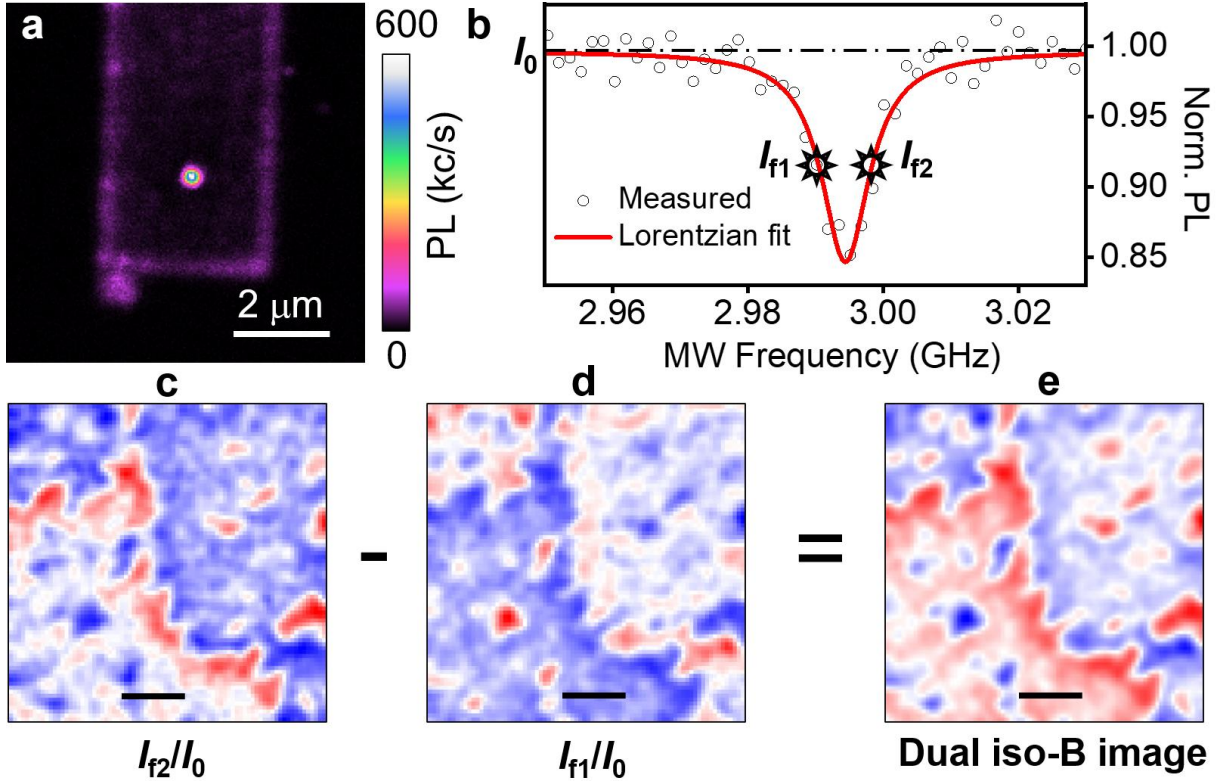

**Figure S2.1.** Protocol for dual iso-B magnetic imaging: (a) Confocal scanning image of the diamond scanning probe. (b) Representative ODMR spectrum of the $|m_S = 0\rangle$ to $|m_S = +1\rangle$ NV spin transition. (c-e) Example of construction of the sign sensitive NV magnetic image (e) by taking the difference of two single MW frequency images at $f_1$ (c) and $f_2$ (d). The scale bar in (c-e) is 200 nm.

The subsequent analysis steps for the retrieval of the mean radius metric are shown in Figure S2.2. Beginning from the raw $B_{str}$ map (Figure S2.2a), the reconstruction of the magnetic moment density component (Fig. S2.2b), $\sigma_z$, is performed based on the geometry of the NV sensor, as outlined in the references [26] and 28]. The $Cr_2O_3$ film is modeled as two planes of spins with out-of-plane magnetic moment and opposite orientation, such that the magnetization $\mathbf{m} = \sigma_z(x, y)[\delta(z) - \delta(z + t)]\hat{z}$, where $\delta$ is the Dirac delta function. The Hanning window defines the region in reciprocal space which is suppressed due to the resolution of the microscope, which is directly proportional to the distance between the NV tip and the surface. The determination of the distinct populations of either up or down surface magnetization enables the binarization of image in Figure 2.2c, which simplifies the calculation of antiferromagnetic (AFM) domain size. In the case of asymmetrically distributed magnetization (B:$Cr_2O_3$), the binarization threshold is defined at the center between the individual gaussian distributions of the bimodal fit. For each of the two populations, the AFM domains are broken up into distinct groups of pixels (see Figure 2.2d), defined by any orthogonal discontinuities. The mean radius, $R_m$, is defined as the mean distance from the domain geometric center to its boundary. Upon another level of averaging, it is a metric by which the information regarding a population of amorphous domains can be compressed to a single value.

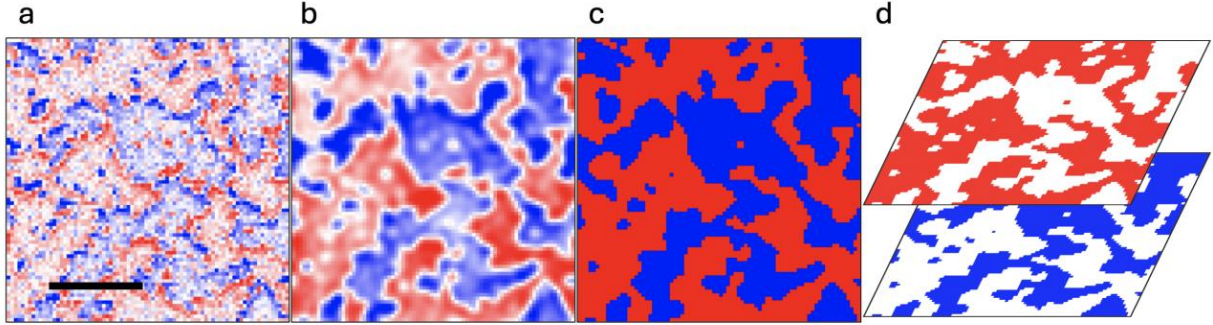

**Figure S2.2.** Protocol for estimation of $\bar{R}_m$ from stray field imaging. (a) $B_{str}$ (raw data) map of 50 nm thin $Cr_2O_3$ film. The scale bar is 1 µm. (b) Reconstructed $\sigma_z$ map from $B_{str}$ map in (a). (c) Binarization of the AFM domain populations from $\sigma_z$ map in (b). (d) Separated order (sign) populations used for the calculation of $\bar{R}_m$ based on grain analysis method in Gwyddion.

## Supplementary Note 3: Fabrication and characterization of capacitors and Hall bar devices made on 200 nm thick B:Cr$_2$O$_3$ films

The hysteresis loops $V_{xy}$ versus $V_G$ were measured at a temperature of 300 K in zero applied magnetic field. The voltage $V_G$ is applied as a quasistatic pulse between top and bottom electrode of the device and removed prior to probing the transverse voltage $V_{xy}$. For acquiring Hall voltage, a readout current of 5 µA is applied and subsequently 100 measurements of transverse voltage $V_{xy}$ are acquired. The sharp transitions at the coercive voltages of about + 7 V and -10 V resemble deterministic switching between distinct AFM states. As detailed in a previously published report,[16] the magnetic field has virtually no effect on the voltage-controlled transition and persists at higher temperature of 370 K and above. These measurements are taken by sourcing the current through a Keithley source meter 6221 and measuring the transverse voltage via Keithley K2182 Nanovoltmeter.

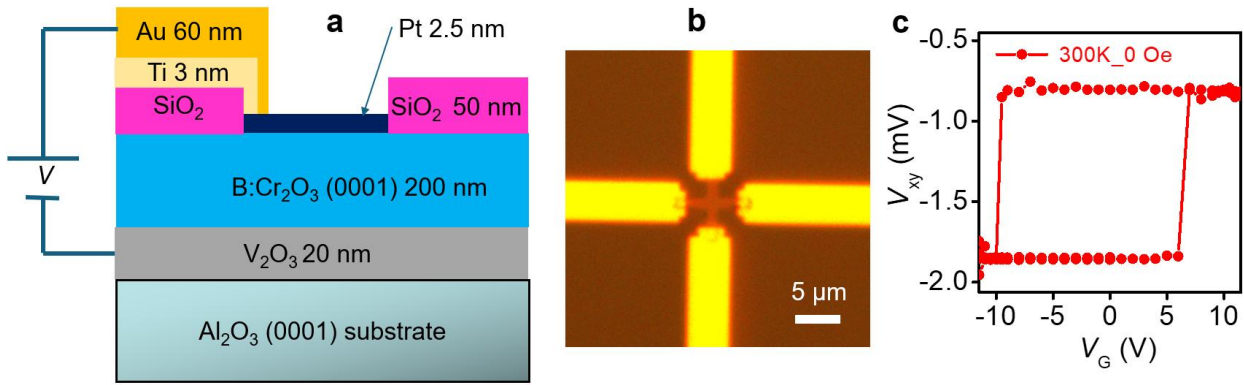

**Figure S3.1.** (a) A schematic showing hybrid structure and device made of voltage switching on B-Cr$_2$O$_3$ films. (b) Optical image of Hall bar devices on B-Cr$_2$O$_3$ (200 nm thick) film. (c) $V_{xy}$ versus $V_G$ hysteresis loops measured at a temperature of 300 K and $B_{app}$ = 0 Oe.

To switch regions locally with the application of electric field/voltage, top electrodes were milled from a continuous Pt layer using focused ion beam (FIB) which is integrated in FEI Helios Nanolab 660. Voltage pulses of 100 µs were applied by biasing a conductive tip in contact with the top electrode. When scanning with the NV-SPM, topography and PL can be recorded simultaneously, as shown in Figures S3.2a and S3.2b, respectively. Due to the modification of the local density of states near Pt (metal),[30] the NV defect PL is comparatively enhanced near the edge of the milled structure (see Figure. S3.2c). From the topography, the milling depth of 10 nm confirms the electrical discontinuity. Upon the application of the reverse polarity, the device "blew up", creating a conductive pathway and preventing further study.

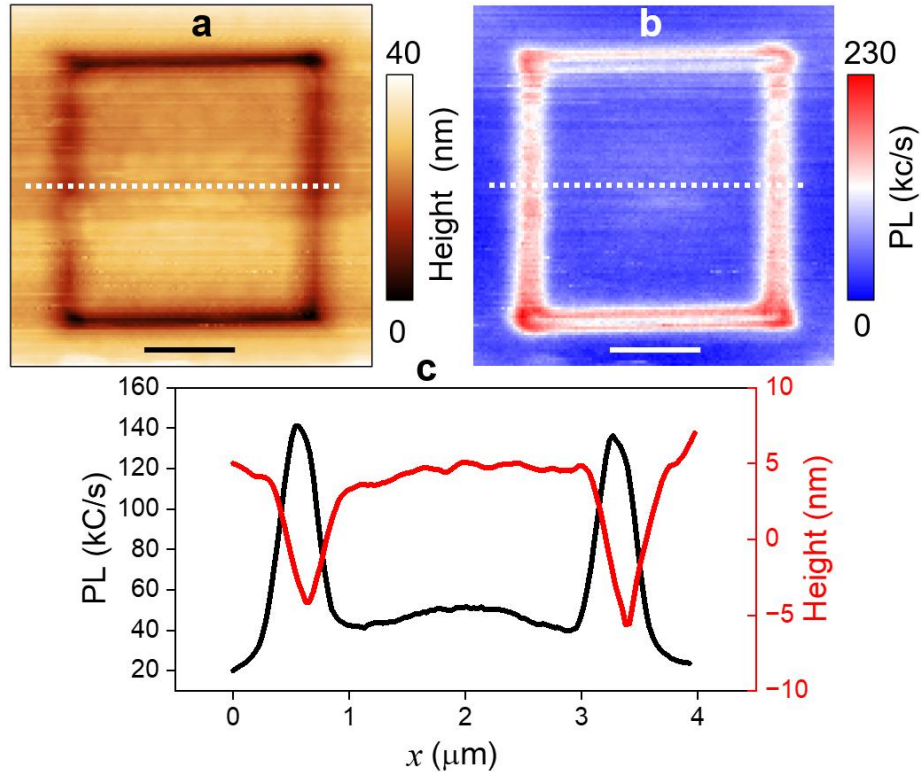

**Figure S3.2.** Topography (a) and PL image (b) of the FIB milled Pt electrode used in electrical field switching of B:$Cr_2O_3$/$V_2O_3$ devices. The scale bar is 1 µm. (c) Line cuts (dashed lines in (a) and (b)) from topography (red) and PL scan (black) which show a milling depth of ~10 nm.